# Defect-Engineered h-BN as a Platform for Single-Atom HER Catalysts: Descriptor Screening Refined by Electrochemical Stability Analysis

Ana S. Dobrota[1], Natalia V. Skorodumova[2], Igor A. Pašti[1,3]*

[1] *University of Belgrade – Faculty of Physical Chemistry, Belgrade, Serbia*

[2] *Applied Physics, Division of Materials Science, Department of Engineering Sciences and Mathematics, Luleå University of Technology, Luleå, Sweden*

[3] *Serbian Academy of Sciences and Arts, Belgrade, Serbia*

*corresponding author

Prof. Igor A. Pašti

*University of Belgrade – Faculty of Physical Chemistry*

*Studentski trg 12-16, 11158 Belgrade, Serbia*

E-mail: igor@ffh.bg.ac.rs

*Phone: +381 11 3336 625*

## Abstract

Defect engineering enables hexagonal boron nitride (h-BN) to act as a platform for stabilizing isolated metal atoms, yet systematic identification of catalytically viable motifs remains limited. Here, density functional theory is used to screen transition and coinage metals anchored at B,

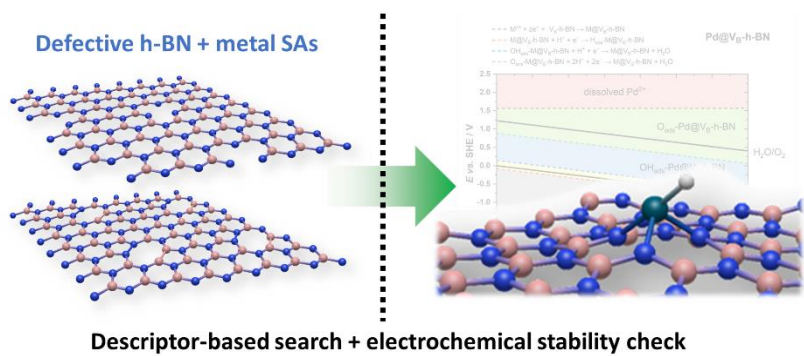

N, and BN vacancies in h-BN for hydrogen evolution reaction (HER) activity. Cohesive-energy benchmarking reveals that B vacancies provide the strongest thermodynamic stabilization of single atoms, while electronic-structure analysis demonstrates vacancy-dependent modulation of conductivity and metal charge state. Hydrogen adsorption free energies identify Cu@$V_N$ and Pd@$V_B$ as near-thermoneutral candidates comparable to Pt(111). However, incorporation of electrochemical stability through Pourbaix analysis significantly refines this selection: Cu@$V_N$ is unstable at low pH and susceptible to $OH_{ads}$ poisoning, whereas Pd@$V_B$ remains stable and catalytically accessible across a broad potential-pH range. These results show that descriptor-based HER screening can generate an expanded pool of candidates, but rigorous electrochemical filtering is essential to identify truly robust systems. The presented multi-step strategy provides a general framework for rational discovery of single-atom catalysts on defect-engineered 2D supports.

## 1. Introduction

Efficient electrocatalysts for the hydrogen evolution reaction (HER) are crucial to sustainable hydrogen production, yet they traditionally rely on scarce noble metals like Pt. The high HER activity of Pt has been linked to its near-optimal binding strength for hydrogen intermediates – neither too weak nor too strong – resulting in a hydrogen adsorption free energy, $\Delta G(H_2)$ close to thermoneutral [1]. Single-atom catalysts (SACs) have emerged as a promising strategy to maximize atomic utilization of such precious metals while maintaining high catalytic activity [2]. By anchoring isolated metal atoms on a support, SACs provide well-defined active sites and often exhibit unique electronic properties and improved selectivity compared to nanoparticle catalysts [2]. A central goal in HER catalyst design is to identify SAC systems where the single metal atoms achieve a $\Delta G(H_2)$ near 0 eV, enabling rapid H adsorption/desorption kinetics akin to Pt [1].

Two-dimensional materials have attracted great interest as supports for SACs due to their large surface area and tunable surface chemistry. In particular, hexagonal boron nitride (h-BN), an insulating analogue of graphene, offers a chemically inert basal plane that can be "activated" by introducing point defects. Pristine h-BN is considered highly inert because of its wide bandgap and strong B–N covalent bonds [3,4], which in turn hinders direct chemical functionalization for catalysis. However, defect engineering creates localized reactive sites in the h-BN lattice. Monatomic vacancies (missing a B or N) disrupt the perfect lattice and introduce mid-gap states and unsaturated dangling bonds [3,4], transforming inert h-BN into a defect-rich surface (sometimes termed d-BN) capable of binding foreign species. Indeed, experimental studies have shown that creating vacancies in h-BN (e.g., via cryo-milling) generates free radical sites that can spontaneously anchor metal atoms from solution [4]. The resulting defective BN loaded with isolated metal atoms exhibits significantly improved HER catalytic performance compared to the pristine material [4]. This ability to switch h-BN from a passive insulator to an active catalyst support underscores the importance of vacancy engineering.

Among point defects in h-BN, boron vs. nitrogen vacancies play distinct roles in stabilizing single metal atoms [2,3,5]. A B-vacancy (one boron atom missing) leaves three under-coordinated N atoms that can strongly coordinate an adsorbed metal, whereas an N-vacancy (missing nitrogen) presents three surrounding B atoms, which are less electron-rich. First-principles calculations confirm that metal adatoms bind much more strongly at B-vacancy sites than at N-vacancy sites in h-BN [2,3]. In a computational survey of 30 metals, Sredojević *et al.* found that the most stable SAC configurations correspond to metals trapped at a missing-B site on h-BN, while metal binding at a missing-N site was generally less stable [2]. Datta and Majumder similarly reported that various transition metals prefer B-vacancies over N-vacancies for energetic stabilization (with the notable exception of Au, which showed an opposite preference) [6]. The stronger trapping ability of B-vacancies is attributed to the favorable metal–N coordination environment and partial ionic character, which can better accommodate and bind the positive metal centers [2]. These findings suggest that B-vacancy engineered h-BN can serve as a robust anchoring platform to prevent single atoms from sintering, a common challenge in SACs [6].

Building on these insights, recent studies have explored h-BN-supported single-atom catalysts for various reactions. Theoretical investigations have screened a range of transition metals on h-BN (defected or modified) to identify promising electrocatalysts. For example, Chen *et al.* examined SACs in a sandwich structure (graphene/metal/h-BN) for bifunctional oxygen reduction and evolution reactions, finding that certain metals (Co, Ni, Cu, etc.) embedded in h-BN can achieve optimal OOH/OH binding energies and low overpotentials for ORR/OER [7]. Sredojević *et al.* evaluated SAC stability in oxidative environments and showed that h-BN vacancies effectively trap many single metal atoms even in the presence of $O_2$ [2]. Another work of Sredojević *et al.* [8] addressed theoretically the HER activity of SACs supported by graphene and h-BN, where defects were also considered, outlining several possible candidates for efficient HER. That work focused on the energetics of hydrogen recombination on supported metals and showed that many described SACs have a favorable energy landscape for HER, proceeding essentially *via* the Tafel mechanism. Experimental work likewise demonstrates the versatility of vacancy-rich h-BN support. Lei *et al.* activated bulk h-BN by creating various vacancy defects, then used the defective surface to stabilize single atoms of metals like Pt and Co; these isolated atoms on h-BN delivered enhanced HER activity compared to non-defective h-BN [4]. In another example, single Cu atoms supported on h-BN catalyze the electroreduction of nitrate to ammonia with improved efficiency, highlighting how a normally inert h-BN sheet can be converted into a catalytic platform when decorated with an appropriate metal center [9]. Across these studies, the introduction of metal atoms onto h-BN consistently modifies its electronic structure (often introducing band-gap states or shifting the Fermi level) and creates active sites for adsorption and redox reactions [9]. This growing body of work positions vacancy-engineered h-BN as a flexible 2D support capable of hosting single-atom active sites for hydrogen, oxygen, and nitrogen-involved electrochemical reactions.

In this work, we present a thorough Density Functional Theory (DFT) study examining single-atom adsorption on vacancy-engineered h-BN, with a focus on hydrogen adsorption as a descriptor for HER activity. We consider a series of transition and coinage metal atoms, Ni, Cu, Ru, Rh, Pd, Ag, Ir, Pt, and Au, representative of different d-block groups and spanning a range of hydrogen affinities. As the anchoring sites, boron and nitrogen monovacancies ($V_B$ and $V_N$, respectively) and boron-nitrogen divacancy ($V_{BN}$) in h-BN were considered. Key properties are evaluated at the DFT level, including the adsorption (binding) energy of the metal adatom, the degree of charge transfer and band gap changes induced by metal binding, and $\Delta G(H_2)$ on the supported single atom as an indicator of HER catalytic activity. Besides indicating the most promising HER candidates, we evaluate their stability at the catalyst|vacuum and the catalyst|electrolyte interface, considering electrode potential and pH, using surface Pourbaix plots for the identified HER candidates [10,11]. In this way, we introduce an additional screening filter for novel SAC-based HER catalysts, in addition to descriptor metrics, and critically revisit some of the previously identified h-BN-based catalytic motifs.

## 2. Computational details

Adsorption of selected transition metals (Ni, Ru, Rh, Pd, Ir, Pt) and coinage metals (Cu, Ag, Au) was investigated on monolayer h-BN. The surface was modeled using a 4×4 supercell containing either a pristine lattice, a single boron vacancy, a single nitrogen vacancy, or a BN divacancy. The chosen supercell size ensures sufficient separation between periodic images of both defects and adsorbates, thereby avoiding spurious interactions arising from periodic boundary conditions. A vacuum region of 20 Å was introduced perpendicular to the surface to eliminate spurious interlayer interactions. All calculations were performed within the framework of density functional theory (DFT) using the Vienna *Ab initio* Simulation Package (VASP) [12–15]. The generalized gradient approximation (GGA) in the Perdew-Burke-Ernzerhof (PBE) parametrization was employed for the exchange-correlation functional [16]. Core-valence interactions were treated using the projector augmented-wave (PAW) method [17,18], with standard PAW potentials supplied with the VASP distribution. Long-range dispersion interactions were accounted for using the DFT-D3 correction of Grimme [19]. The plane-wave kinetic-energy cutoff was set to 520 eV. Electronic occupations were treated using Gaussian smearing with a width of $\sigma = 0.025$ eV. Brillouin-zone sampling was carried out using a Γ-centered Monkhorst-Pack grid of 10×10×1 k-points. Spin polarization was included in all calculations. Structural relaxations were performed until the Hellmann-Feynman forces on all atoms were smaller than $10^{-2}$ eV Å$^{-1}$, corresponding to a total-energy convergence better than 0.01 meV. Visualization of optimized geometries was performed using VESTA [20] and Avogadro [21]. Densities of states (DOS) were extracted and analyzed using the sumo toolkit for VASP [22]. Band gaps were determined using VASPKIT [23].

The binding energy of metal M onto defective h-BN was calculated as:

$$E_b(M) = E_{V_X\text{-h-BN+M}} - E_{V_X\text{-h-BN}} - E_M \qquad (1)$$

where $E_{V_X\text{-h-BN+M}}$ is the total energy of the defective h-BN containing a $V_X$ vacancy (X = B, N, or BN) with the metal atom M adsorbed, $E_{V_X\text{-h-BN}}$ is the energy of the corresponding clean defective h-BN slab, and $E_M$ is the energy of the isolated metal atom. Negative values of $E_b$ indicate exothermic adsorption. The binding energies of other adsorbates at metal sites on metal-functionalized defective h-BN (hydrogen, oxygen, hydroxyl group) were calculated in an analogous way and then converted to the Gibbs free energy of adsorption. In particular, the Gibbs free energy for $H_2$ adsorption $\Delta G(H_2)$, was referred to molecular $H_2$ in the gas phase.

Vibrational analysis was performed using a second-order finite-difference approach with atomic displacements of 0.015 Å in all three Cartesian directions. Constrained dynamics was employed such that vibrational frequencies were calculated by displacing the adsorbate and the active single-atom catalytic site, including atoms up to the second-nearest neighbors of the metal center. This approach significantly reduces computational cost while maintaining satisfactory accuracy; within this approximation, Gibbs free energies are underestimated by approximately 0.03-0.06 eV [24]. Zero-point energy (*ZPE*) corrections and vibrational entropy contributions ($TS_{vib}$) were obtained from the calculated vibrational frequencies.

Electrochemical reaction energetics were evaluated using the Computational Hydrogen Electrode (CHE) approach [25]. Assuming equilibrium of the hydrogen electrode, $H^+ + e^- \leftrightarrow H_{2(g)}$, the electrochemical potential of ($H^+ + e^-$) was equated to that of gaseous $H_2$ at pH = 0. For each intermediate *i*, the chemical potential was calculated as:

$$\mu_i = E_{tot} + ZPE - TS_{vib}, \qquad (2)$$

Electric-field effects acting through surface dipole moments were neglected, as justified in Ref. [26]. The chemical potential of liquid water at 298 K was derived from the vapor chemical potential at the corresponding equilibrium vapor pressure. Equilibrium potentials for electrochemical reactions were obtained by treating the reaction as a cathode in a hypothetical galvanic cell with the standard hydrogen electrode (SHE) as the anode, following established procedures [27]. Gibbs free-energy changes ($\Delta G$) were calculated accordingly, and electrode potentials were obtained by dividing $\Delta G$ by the number of electrons transferred. Since the SHE anode potential is defined as 0 V, the resulting values correspond directly to standard electrode potentials. Metal dissolution potentials from vacancy sites were calculated following the methodology described in Ref. [28], considering a hypothetical galvanic cell consisting of a bulk metal electrode and an $M@V_X$-h-BN electrode. For the construction of the Pourbaix diagram, the activity of the metal ion $M^{z+}$ was set to $1\times10^{-8}$ mol dm$^{-3}$. Changing the metal-ion concentration by one order of magnitude shifts the equilibrium potential vertically by 0.059 V at room temperature.

The thermal stability of selected adsorption configurations was further assessed using *ab initio* molecular dynamics simulations at 298 K, employing a Nosé-Hoover thermostat [29,30]. A time step of 1 fs was used, and simulations were run for 1000 fs.

## 3. Results
### 3.1. Vacancies in h-BN

The optimized structures and the densities of states for the three considered types of vacancies in h-BN are presented in **Figure 1**. Upon vacancy formation, the h-BN sheet remains flat, with small structural changes around the vacancy sites, while the electronic structure and magnetic properties change drastically. The $V_N$ formation induces defect states around the Fermi level, and the magnetization appears (1 $\mu_B$). The rise of the defect states results in the reduction of the h-BN band gap (4.70 eV) to 0.51 eV. Similarly, the formation of $V_B$ results in empty states well localized on nitrogen atoms surrounding the vacancy. A total magnetization of 3 $\mu_B$ is observed, and the band gap is 0.11 eV. Finally, the induction of $V_{BN}$ results in a non-magnetic system, with a rather wide band gap of 3.04 eV.

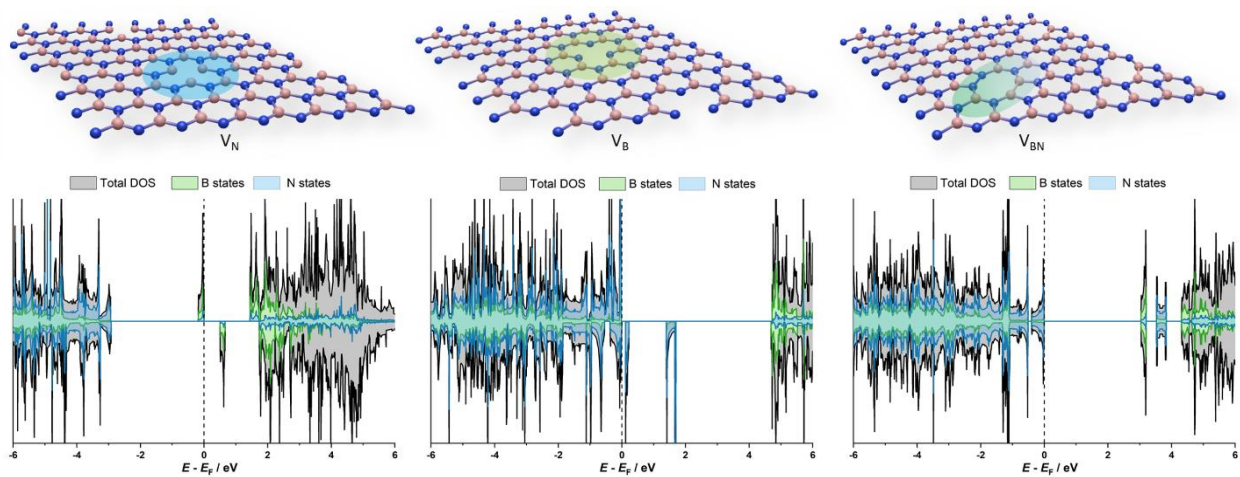

**Figure 1.** Optimized structures (up) and DOSes (down) of h-BN with: nitrogen vacancy ($V_N$, left), boron vacancy ($V_B$, middle), and BN divacancy ($V_{BN}$, right) (blue atoms – nitrogen, pink atoms – boron).

These structural and electronic modifications directly redefine the chemical landscape available for metal adsorption. The appearance of vacancy-induced states near the Fermi level introduces electronically accessible orbitals that can participate in metal–support coupling, in contrast to pristine h-BN, where the large band gap limits such interaction. In the case of $V_N$, the presence of partially occupied defect states and a finite magnetic moment indicates the availability of localized spin-polarized states that can interact with metal d orbitals. For $V_B$, the combination of strongly localized empty states on the neighboring nitrogen atoms, substantial band-gap narrowing, and a higher total magnetization reflects an even more pronounced electronic perturbation of the lattice. Such features imply a highly modified local electronic environment at the vacancy site compared to the inert basal plane. In contrast, the $V_{BN}$ defect restores a relatively wide band gap and a non-magnetic ground state, indicating a more electronically compensated configuration. Thus, although all vacancy types preserve the overall planarity of the sheet, they generate distinctly different local electronic structures, which provide the fundamental framework governing metal-support interaction strength, orbital hybridization, and magnetic coupling upon adsorption.

### 3.2. Metal anchoring to vacancy sites

Thermodynamic stability of the chosen transition and coinage metals at h-BN vacancy sites was assessed by comparing their binding energies to the corresponding elemental cohesive energies, as shown in **Figure 2**. The upper-left panel directly compares the metal binding energy with the cohesive energy ($E_{coh}$). The dashed line ($|E_b| = E_{coh}$) represents the thermodynamic threshold separating stabilization of isolated atoms (below the line) from clustering preference (above the line). This representation provides a universal, metal-independent criterion for assessing single-atom stability. The results compare well with the previously published data by Sredojevic *et al.* [2] for the cases of $V_B$ and $V_N$.

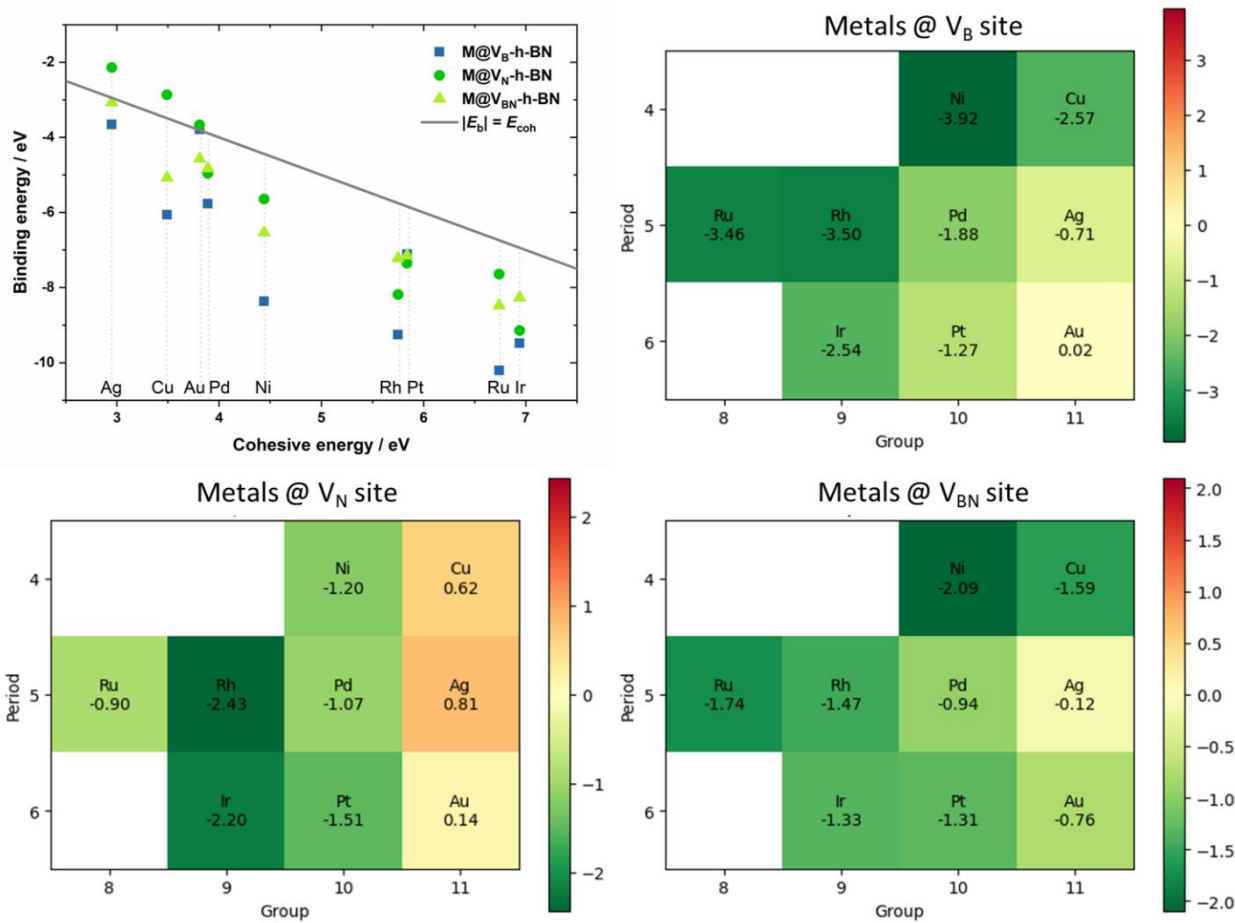

**Figure 2.** Upper left – thermodynamic stabilization of transition and coinage metals at vacancy sites of h-BN: binding energy of metal atoms at $V_B$ (blue squares), $V_N$ (dark green circles), and $V_{BN}$ (light green triangles) plotted against their elemental cohesive energies, which separates stabilized single atoms (below the line) from configurations favoring clustering (above the line). Heat maps of binding energies at the vacancy sites (upper right – $V_B$, lower left – $V_N$, lower right – $V_{BN}$), arranged by periodic group and period, and referred to the metal cohesive energy.

A clear hierarchy of vacancy performance emerges. The $V_B$ site exhibits the strongest stabilization across the metal series. Ru, Rh, and Ni lie well below the parity line, indicating that binding to $V_B$ overcomes the thermodynamic driving force toward bulk aggregation. Iridium also shows strong stabilization, while Pd lies close to the threshold. In contrast, the coinage metals (Cu, Ag, Au) approach or exceed the parity condition, indicating substantially weaker anchoring at $V_B$ than in early and mid-transition metals. The heat maps further quantify these trends by arranging binding energies by group and period while implicitly referencing cohesive energies through the parity comparison. At $V_B$, binding strength systematically decreases from Groups 8-9 toward Group 11 within a given period. For example, in the 4d row, Ru and Rh bind considerably more strongly than Pd and Ag. Similar behavior is observed in the 5d series, where Ir shows stronger stabilization than Pt and Au. At the $V_N$ site, stabilization is significantly reduced for all metals compared to $V_B$. Rh and Ir remain clearly below the cohesive energy threshold, whereas Ni and Pd lie closer to marginal stability. Coinage metals display weak or even positive binding energies, confirming insufficient anchoring against aggregation. The $V_{BN}$ defect exhibits intermediate behavior. However, it is interesting to observe that this is the only type of vacancy that stabilizes all the metals below their cohesive energy. The cohesive-energy benchmarking demonstrates that single-atom stabilization on h-BN is strongly defect-dependent. Among the investigated sites, $V_B$ provides the most favorable thermodynamic environment for anchoring isolated transition metals, while $V_N$ is comparatively less effective. On the other hand, $V_{BN}$ is expected to effectively stabilize all the considered metals (transition and coinage ones) in a single-atom form.

Metal anchoring at different types of vacancies reshapes the electronic structure of defective h-BN and modulates the charge state of the isolated metal atoms, as shown in **Figure 3**. These two factors are directly relevant to the electrochemical functionality of any material. For electrochemical applications, electronic conductivity is essential, and the calculated band gaps clearly distinguish the vacancy types in this respect. At the $V_B$ site, several systems exhibit complete or near-complete band-gap closure (e.g., Ni, Pd, and Pt), while others retain only small residual gaps. This indicates that anchoring at $V_B$ frequently transforms the wide-gap insulating h-BN host into a nearly metallic or highly conductive material. In contrast, the $V_N$ site generally preserves finite band gaps, although the band gap is substantially narrowed relative to pristine h-BN. The $V_{BN}$ defect leads to intermediate behavior, with moderate band-gap reductions but no full closure. The Bader charge analysis reveals systematic differences in the electronic character of the anchored metals. At $V_B$, metals carry positive partial charges, indicating net electron transfer from the metal to the surrounding nitrogen atoms. The magnitude of this charge is particularly pronounced for Ni and Cu, and remains significant across the transition-metal series. At $V_N$, the trend reverses: metals generally acquire negative partial charges, reflecting electron donation from the boron-coordinated vacancy environment to the metal center. This inversion of charge polarity between $V_B$ and $V_N$ emphasizes the fundamentally different electronic nature of the two defects. The $V_{BN}$ site, again, exhibits intermediate behavior, with smaller absolute charge magnitudes and more modest polarization of the metal centers, both positive and negative.

Importantly, the magnitude of charge transfer correlates qualitatively with band-gap reduction: stronger charge redistribution at $V_B$ accompanies more pronounced electronic delocalization, whereas weaker charge polarization at $V_{BN}$ coincides with retention of finite gaps. From an electrochemical perspective, $V_B$-embedded metals combine two advantageous features, enhanced conductivity and significant modulation of the metal electronic state, both of which are critical for catalytic charge transfer processes. Thus, the vacancy type not only governs thermodynamic stabilization but also controls the electronic conductivity and effective oxidation state of the anchored metal centers.

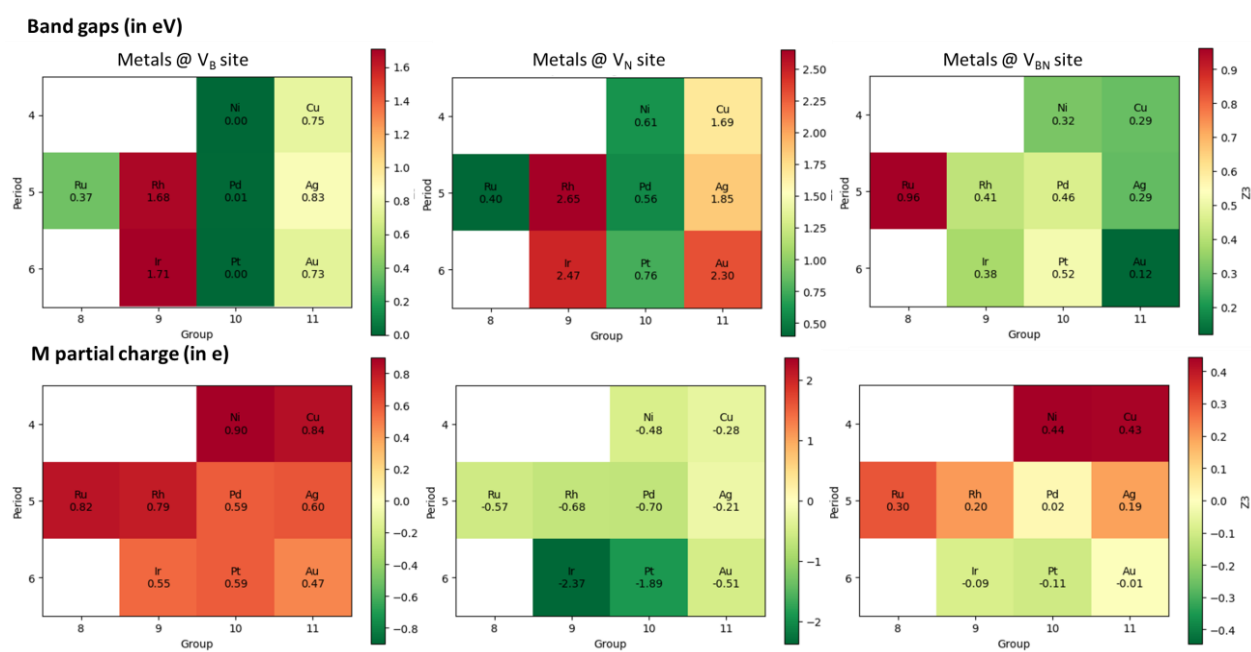

**Figure 3.** Bandgaps and partial (Bader) charges for metal atoms anchored at the vacancy sites in h-BN.

### 3.3. Hydrogen binding at the SAs and HER activity

The Gibbs free energies of hydrogen adsorption reveal a strong vacancy-dependent modulation of the intrinsic hydrogen-binding properties of the anchored metals (**Table 1**). As expected for an efficient hydrogen evolution catalyst, values close to thermoneutrality are desirable. At the B-vacancy site, a broad spread of hydrogen binding strengths is observed. Pd (0.083 eV) and Ir (0.218 eV) lie closest to thermoneutral adsorption, whereas Pt (−0.648 eV) and Ru (−0.550 eV) bind hydrogen considerably too strongly. In contrast, Ag, Cu, and Ni exhibit overly weak hydrogen adsorption, reflected by large positive ΔG(H$_2$) values. The situation changes markedly at the N-vacancy site, where Ni (−0.695 eV), Pd (−0.506 eV), and Pt (−0.942 eV) bind hydrogen more strongly than at V$_B$, while Cu approaches thermoneutral behavior (0.048 eV). However, Rh and Ir exhibit significantly positive ΔG(H$_2$) values at V$_N$, indicating weak hydrogen stabilization. The BN divacancy produces yet another regime: several metals (Rh −1.135 eV, Ir −1.268 eV, Ni −0.828 eV, Pt −0.848 eV) display very strong hydrogen binding, whereas Cu (−0.387 eV) and Au (−0.331 eV) show moderate adsorption strengths closer to optimal values.

**Table 1.** Gibbs energies of hydrogen adsorption at metal atoms embedded in vacancies in h-BN.

| Metal (M) | ΔG(H$_2$) / eV | | | |
|---|---|---|---|---|
| | M@V$_B$-h-BN | M@V$_N$-h-BN | M@V$_{BN}$-h-BN | M(hkl)* |
| Ag | 1.321 | 0.478 | −0.652 | 0.51[a] |
| Au | 0.589 | 0.446 | −0.331 | 0.45[a] |
| Cu | 1.397 | 0.048 | −0.387 | 0.19[a] |
| Ir | 0.218 | 1.004 | −1.268 | 0.03[a] |
| Ni | 0.842 | -0.695 | −0.828 | −0.27[a] |
| Pd | 0.083 | −0.506 | −0.539 | −0.14[a] |
| Pt | −0.648 | −0.942 | −0.848 | −0.09[a] |
| Rh | 0.657 | 1.273 | -1.135 | −0.10[a] |
| Ru | −0.550 | 0.815 | −0.215 | 0.133[b] |

*(111) for Ag, Au, Cu, Ir, Ni, Pd, Pt, and Rh, (0001) for Ru, [a] ref. [31], [b] ref. [32]

Comparison with the corresponding low-index metal surfaces highlights the substantial electronic reconfiguration caused by anchoring. For example, Pt(111), which is nearly thermoneutral (−0.09 eV), becomes significantly more prone to hydrogen binding when a Pt single atom is anchored at any vacancy site in h-BN. Similarly, Ni and Pd show notable deviations from their surface values depending on the defect type. In some cases, anchoring weakens hydrogen adsorption compared to the metal surface (e.g., Cu at V$_N$ *vs.* Cu(111)), while in others it greatly strengthens it (e.g., Cu and Ir at V$_{BN}$). These shifts indicate that vacancy anchoring does not simply mirror the metal's intrinsic surface chemistry but creates a distinct electronic state that alters hydrogen affinity. The data demonstrate that both the magnitude and the direction of ΔG(H$_2$) shifts are strongly dependent on the type of vacancy, underscoring the critical role of defect engineering in modulating hydrogen-binding thermodynamics.

The trends in ΔG(H$_2$) can be rationalized in connection with the charge redistribution shown in **Fig. 3**. At the V$_B$ site, metals generally carry positive Bader charges, indicating electron transfer from the metal to the surrounding nitrogen atoms. In this electron-depleted state, hydrogen adsorption is often weakened, which is consistent with the predominantly positive ΔG(H$_2$) values observed for Ag, Au, Cu, Ni, and Rh. In contrast, at the V$_N$ site, the metals acquire negative partial charges due to electron donation from the boron-coordinated environment. This electron enrichment is accompanied, in several cases (e.g., Ni, Pd, Pt), by more negative ΔG(H$_2$) values than those of V$_B$, indicating stronger hydrogen binding. The $V_{BN}$ defect exhibits smaller charge magnitudes and, correspondingly, intermediate hydrogen adsorption strengths. While the correlation is not strictly quantitative for all metals, the overall trend indicates that vacancy-induced charge redistribution contributes to systematic shifts in hydrogen-binding thermodynamics, with electron-rich metal centers generally exhibiting stronger hydrogen adsorption than electron-deficient ones.

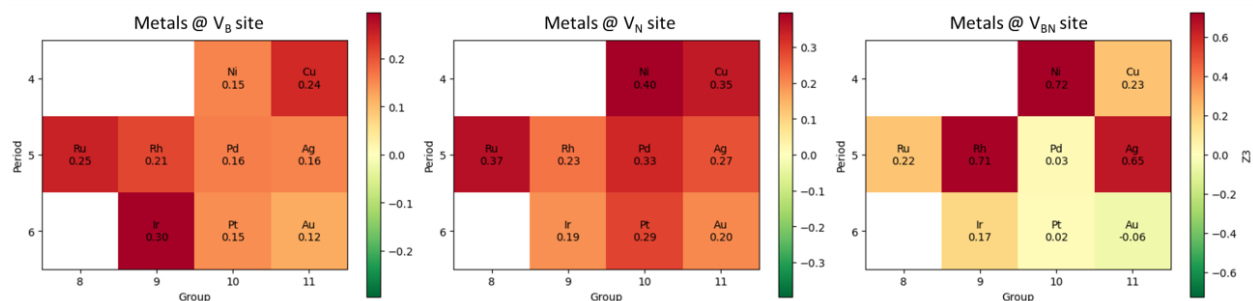

**Figure 4.** Partial (Bader) charges on $H_{ads}$ atoms bound to metals anchored at different vacancy sites in h-BN.

The Bader charges on adsorbed hydrogen (**Figure 4**) show that H carries a positive partial charge in all configurations, except for Au at the $V_{BN}$ site, indicating dominant electron transfer to the metal-vacancy complex. The magnitude of this charge depends on the vacancy type. At $V_B$, hydrogen charges are moderate and relatively uniform across the series, consistent with the generally weak to near-thermoneutral hydrogen adsorption observed at this site. At $V_N$, hydrogen charges are systematically larger for several metals, which aligns with stronger hydrogen binding. The $V_{BN}$ site exhibits a wider spread: systems with higher hydrogen charge (e.g., Ni, Rh, and Ag) correspond to strongly negative ΔG(H₂), whereas lower hydrogen charge is associated with weaker adsorption. Although not strictly quantitative, the data suggest a qualitative link between increased M–H bond polarization (i.e., a higher positive charge on H) and stronger hydrogen stabilization, reflecting vacancy-dependent electronic modulation of hydrogen binding.

Based on the hydrogen adsorption thermodynamics, Cu@$V_N$ and Pd@$V_B$ emerge as the two most promising HER candidates within the investigated set. Both systems exhibit ΔG(H₂) values closest to thermoneutral and comparable in magnitude to Pt(111), and this is consistently reflected in their reaction free-energy profiles as well as in their positions near the apex of the HER volcano plot (**Figure 5**). In this sense, both configurations satisfy the Sabatier criterion, indicating balanced hydrogen adsorption strength that is neither too weak nor too strong. However, their electronic structures differ noticeably (**Figure 5**). The projected DOS for Cu@$V_N$ retains a relatively large band gap (>1.5 eV), consistent with the semiconducting character of the $V_N$-based system even prior to hydrogen adsorption. In contrast, Pd@$V_B$ exhibits a substantially smaller band gap after H adsorption, although still slightly larger than in the H-free Pd@$V_B$ system, which is nearly metallic. This difference suggests that Pd@$V_B$ may offer more favorable intrinsic electronic conductivity, whereas Cu@$V_N$ remains electronically more insulating.

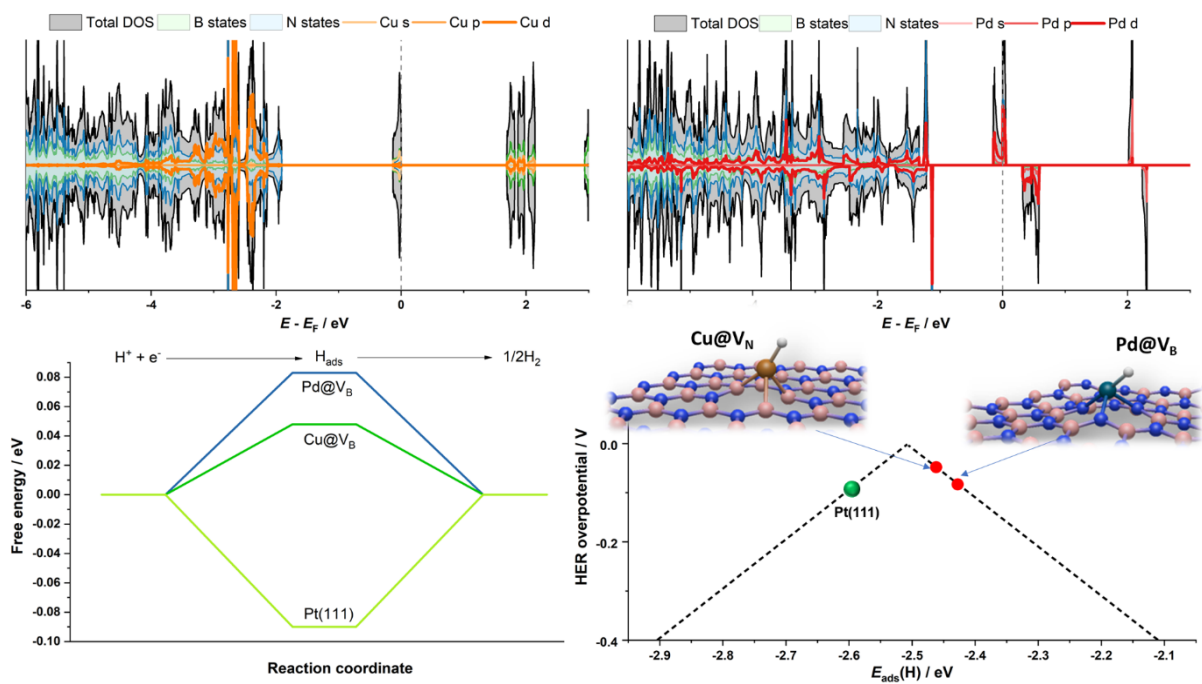

**Figure 5.** Top left: total and projected densities of states for Cu@$V_N$ after hydrogen adsorption, showing contributions from B, N, and Cu (s, p, d) states. Top right: total and projected DOS for Pd@$V_B$ after hydrogen adsorption. The Fermi level is set to 0 eV. Bottom left: reaction free-energy profiles for hydrogen adsorption (H$^+$ + e$^-$ → H$_{ads}$ → ½H$_2$) comparing Cu@$V_N$, Pd@$V_B$, and Pt(111) as a reference. Bottom right: position of Cu@$V_N$ and Pd@$V_B$ on the HER volcano plot as a function of hydrogen adsorption energy, with Pt(111) indicated for comparison. The proximity of both vacancy-anchored systems to the volcano apex and their near-thermoneutral ΔG(H$_2$) values indicate that they are promising HER motifs despite residual band gaps.

From a screening perspective, however, the descriptor-based evaluation focuses on the intrinsic thermodynamics of hydrogen binding at the active site rather than the macroscopic conductivity of the composite electrode. In practical electrochemical applications, finite band gaps can be effectively addressed by incorporating conductive additives or by forming composite structures, thereby facilitating efficient charge transport to and from the catalytic center. Therefore, if the screening ends at the level of ΔG(H$_2$), reaction profiles, and volcano placement, both Cu@$V_N$ and Pd@$V_B$ can be considered viable HER-relevant motifs, with the understanding that their different electronic gaps may influence electrode design strategies rather than their intrinsic catalytic thermodynamics. However, we need to step back and examine the SAC formation step, which shows that copper binding to the $V_N$ site does not exceed the cohesive energy of metallic Cu. As a result, agglomeration could occur. We discuss the stability issue in the next section and extend this consideration to the actual state of the SA-center under electrochemical conditions.

### 3.4. Stability of the HER candidates

The strength of the metal-vacancy interaction is depicted in **Figure 6** (top row) by comparing the adsorption energies of Cu@$V_N$ and Pd@$V_B$ relative to their binding on pristine h-BN. In both cases, the vacancy provides a substantially deeper potential well, with stabilization energies of 3.26 eV for Cu@$V_N$ and 6.94 eV for Pd@$V_B$ compared to adsorption on the defect-free sheet. These large energy differences indicate that the vacancy sites act as strong trapping centers, creating potential wells that are far deeper than the thermal energy available at room temperature ($k_BT \approx 0.026$ eV). Consequently, migration from the vacancy site to the pristine basal plane is thermodynamically unfavorable under ambient conditions. Notably, even though the binding of Cu at $V_N$ does not exceed its cohesive energy in the parity analysis, the pronounced depth of the local potential well suggests kinetic stabilization of the single-atom configuration. This interpretation is supported by the *ab initio* MD simulations (**Figure 6**, bottom row), which show stable total-energy fluctuations without indications of metal detachment or surface diffusion over the simulated time scale at 298 K. Together, these results demonstrate that both Cu@$V_N$ and Pd@$V_B$ are strongly confined at their respective vacancy sites, reinforcing their viability as structurally stable single-atom motifs. Thus, stopping the analysis with the confirmation of stability at the catalyst|vacuum interface would leave these two candidates as the viable ones for experimental verification.

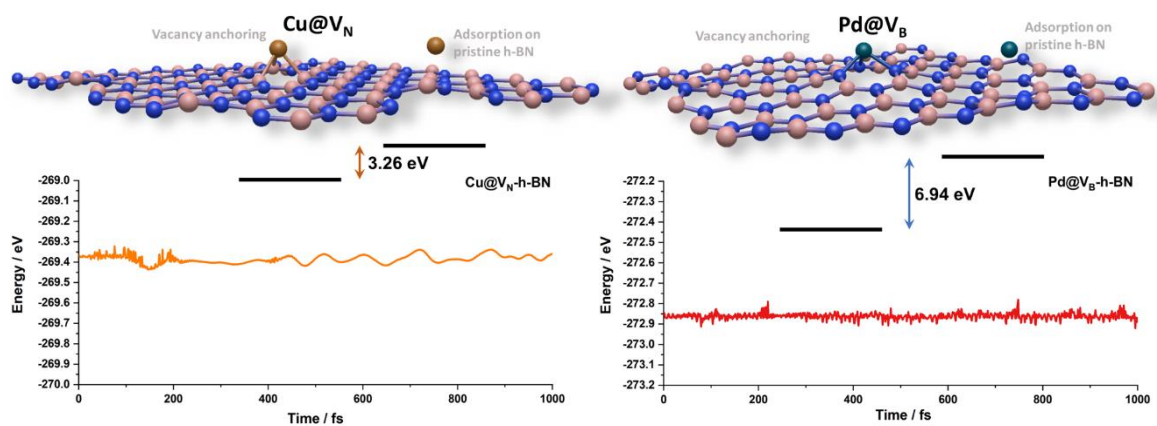

**Figure 6.** Thermodynamic considerations of the stability of HER catalyst candidates. Top: schematic depiction highlighting the energy difference between vacancy anchoring and adsorption on pristine h-BN (3.26 eV for Cu@$V_N$, and 6.94 eV for Pd@$V_B$), illustrating the depth of the vacancy-induced potential well. Bottom row: molecular dynamics simulations at 298 K showing stable total-energy fluctuations over 1 ps, confirming the kinetic stability of the single-atom configurations.

In the last step, we analyze the stability of Cu@$V_N$ and Pd@$V_B$ systems under electrochemical conditions, using the Pourbaix plots as described previously in ref. [10,11]. Note that we do not consider possible solid phase formation upon the dissolution of Cu and Pd, as later described by Di Liberto *et al*. [33], as we are primarily interested in the cathodic region where HER is operative. The final stability filter is derived from Pourbaix analysis under electrochemical conditions (**Figure 7**). For Cu@$V_N$, the stability window is limited. At low pH, the dominant region corresponds to dissolved $Cu^{2+}$, indicating thermodynamic instability of the anchored Cu species in acidic environments. At higher pH values, although dissolution is suppressed, the clean Cu@$V_N$ surface is not the most stable state. Instead, $OH_{ads}$ covered configurations dominate over a broad potential range. This implies that under alkaline or near-neutral conditions, the active site would be partially or fully hydroxylated, thereby modifying or blocking the catalytic center and altering its hydrogen-adsorption properties. Consequently, even though Cu@$V_N$ exhibits near-thermoneutral $\Delta G(H_2)$ and favorable volcano positioning, its electrochemical stability window for HER-relevant operation is restricted and subject to surface poisoning by hydroxyl species.

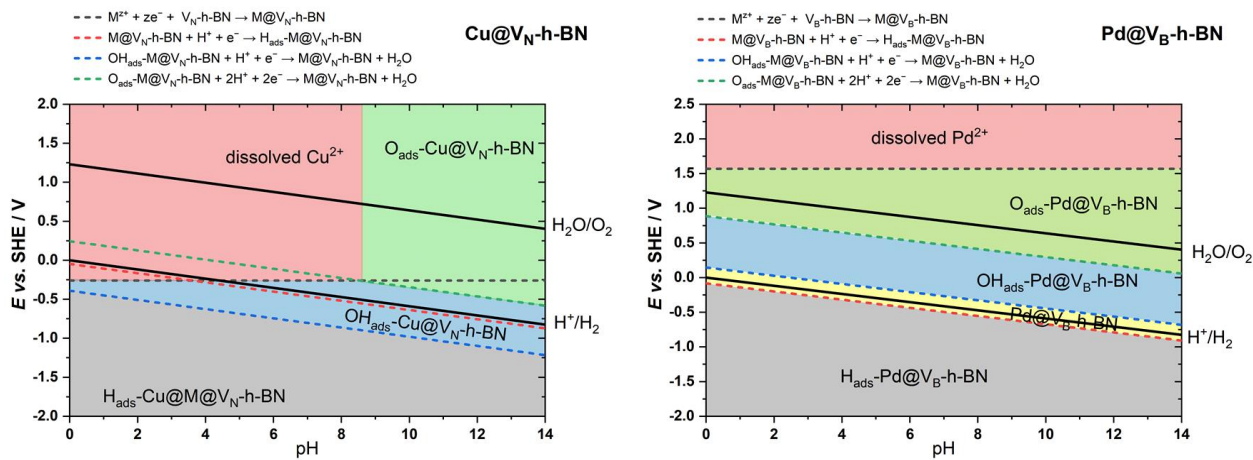

**Figure 7.** Pourbaix plots for Cu at the $V_N$ site (left) and Pd at the $V_B$ site (right) in the h-BN sheet.

In contrast, Pd@$V_B$ demonstrates markedly superior electrochemical robustness. The Pourbaix diagram shows no dissolution region within the HER-relevant potential window, and a distinct stability domain exists for the clean Pd@$V_B$ configuration. Upon moving to more cathodic potentials, $H_{ads}$ becomes the stable surface species, directly aligning with the onset of the hydrogen evolution reaction. Importantly, no extensive $OH_{ads}$-dominated region interferes with HER operation. Combined with its near-thermoneutral $\Delta G(H_2)$, favorable volcano placement, and relatively small band gap upon hydrogen adsorption, Pd@$V_B$ satisfies all screening criteria: thermodynamic hydrogen binding, structural stability, electrochemical resilience, and absence of competitive surface poisoning. Given that $\Delta G(H_2)$ serves as a valid HER descriptor across pH values [34], the stability map suggests that Pd@$V_B$ remains operative over a wide pH range. After applying comprehensive multi-step screening, adsorption thermodynamics, cohesive-energy benchmarking, kinetic stability, and electrochemical Pourbaix analysis, Pd@$V_B$ emerges as the only viable HER candidate among the 27 SAC systems investigated. To reinforce the importance of the assessment of the electrochemical stability after the descriptor-based search, we note that the here-presented Pourbaix plot for Pd@$V_B$ agrees perfectly with the one presented in ref. [35]. However, in the mentioned work, the authors did not consider the formation of Pd-$H_{ads}$ species, which happens just below 0 V vs. RHE. Considering that the mentioned work was focused on the development of $CO_2$ reduction reaction catalysts, the formation of $H_{ads}$ is extremely important, as our results suggest that the $CO_2$ reduction on Pd@$V_B$ shall not take place on bare Pd sites, but on the Pd single atoms modified by $H_{ads}$.

## 4. Conclusions

In this work, we systematically investigated the stabilization and catalytic relevance of selected transition and coinage metals anchored at vacancy sites in hexagonal boron nitride using a unified descriptor-based DFT framework. By combining adsorption energetics, cohesive-energy benchmarking, electronic-structure analysis, hydrogen-adsorption thermodynamics, molecular-dynamics stability tests, and electrochemical Pourbaix evaluation, we established a multi-level screening protocol to identify viable single-atom HER moieties. Cohesive-energy comparisons revealed that vacancy type decisively governs thermodynamic stabilization, with $V_B$ sites providing the strongest anchoring environment. Electronic-structure analysis showed that vacancy-induced states enable strong metal–support coupling and modulate conductivity, while Bader charge trends demonstrated systematic control of hydrogen affinity through vacancy-dependent charge redistribution. Descriptor-based screening using $\Delta G(H_2)$ identified Cu@$V_N$ and Pd@$V_B$ as promising HER candidates, both located near the apex of the volcano and comparable to Pt(111). However, inclusion of electrochemical stability through Pourbaix analysis fundamentally refined this selection: Cu@$V_N$ is limited by dissolution in acidic media and OH$_{ads}$ poisoning at higher pH, whereas Pd@V$_B$ remains stable, unpoisoned, and catalytically accessible across a broad pH window. Our results demonstrate that descriptor-based searches alone can overestimate the number of viable catalytic motifs. Only when thermodynamic hydrogen binding is complemented by cohesive-energy and electrochemical stability filters does a truly robust candidate emerge. Within this integrated framework, Pd@V$_B$ stands out as the most promising HER-active configuration. More broadly, the approach presented here provides a transferable strategy for the rational discovery of single-atom catalysts on defect-engineered 2D materials.

## Acknowledgement
A.S.D. and I.A.P. acknowledge the financial support provided by the Serbian Ministry of Science, Technological Development, and Innovations (contract no. 451-03-34/2026-03/200146) and the Serbian Academy of Sciences and Arts (project F-190). The computations and data handling were enabled by resources provided by the National Academic Infrastructure for Supercomputing in Sweden (NAISS) at the National Supercomputer center (NSC) at Linköping University, partially funded by the Swedish Research Council through grant agreements No. NAISS 2024/5-718 and NAISS 2025/5-713.